\documentclass[12pt]{article}

\def\beqra{\begin{eqnarray}}
\def\eeqra{\end{eqnarray}}
\def\beqast{\begin{eqnarray*}}
\def\eeqast{\end{eqnarray*}}
\def\be{\begin{enumerate}}
\def\ee{\end{enumerate}}

\def\beq{\begin{equation}}
\def\eeq{\end{equation}}

\begin{document}



\vspace{24pt}

\begin{center}
{\large{\bf   A Model of Hot-Sector Generations}}

\vspace{36pt}
K. Koike

\vspace{4pt}
Department of Natural Science \\ Kagawa University
\\ Takamatsu, 7608522 Japan

\vspace{30pt}
{\bf Abstract}

\end{center}

\begin{minipage}{4.75in}

Possible existence of ``hot-sector generations" above the well known 3 
generation bound is investigated on the basis of a model of leptons and
quarks, which is based on the Harari and Shupe's one. Our model predicts
the existence of {\bf3 + 1} generations above the ordinary ``cold-sector" 
3 generations. Majorana neutrinos are introduced 
to realize the heavy neutrino masses in hot-sector generations. Properties 
of heavy neutrinos are also discussed.

\end{minipage}

\vfill

\baselineskip=24pt
\pagebreak
\setcounter{page}{1}

\section{Introduction}

The 3 generation structure appearing in low energy region seems to bring 
suggestions concerning to the deeper level of nature. Is the number of 
generations restricted to just 3? 
A certain kind of models predict possible existence of ``hot-sector 
 generations" above the 3 generation structure. In what form can the 
 hot-sector generations exist? This paper is concerning to this problem.
 
 The concept of hot-sector generation had been proposed by Maki\cite{rf:Maki}
in a  consideration of Blokhintsev type\cite{rf:Blokhin} for the meaning of 
presently appearing generation structure. He discussed that the ``standard" 
particle picture will not hold in very high-energy region, and the introduction 
of very heavy particles would upset easily and drastically the standard 
physical features of the fields participating in the low-lying generations 
as a whole, leading the standard model to be almost meaningless. 
Then, the single question ``how many generations are there in nature" 
should be divided into two similar questions as regards to ``the  
cold- and hot-sector generations"\cite{rf:Maki}, respectively.
    
  Thus, the famous decision\cite{rf:Gener3} of the generation number resulting 
from  $Z \rightarrow \overline{\nu_{\ell}}\nu_{\ell}$ experiment should be 
interpreted that the number of ``cold-sector generations" is 
just 3. Then, what reason is  there  behind  this fact. What structure is expected 
for the hot-sector generations?  It should be emphasized that in such energy 
region as hot-sector generations, the dynamics and particle picture should be 
drastically altered from ordinary field theory, then it is meaningful to  
investigate a simple model to realize the characteristic features of phenomena. 
These circumstances should be compared with early stage of quark model  
\cite{rf:quark1,rf:quark2}. 
We will suppose that
our model is concerning to the sub-structure, which suggests a guiding principle to 
build a model in the framework of GUTs structure.\cite{rf:quark0}

\section{Schematical model of generations}

 For definiteness, we will refine and summarize the essence of our previous work on 
generation structure\cite{rf:KK}.
Our model of generations is based on the 
rishon model\cite{rf:rishon} of Harari and Shupe, where all leptons and quarks
are 3 body system of rishons T with charge 1/3 and V with charge 0, and freedom
of color are realized by their configuration.

 In our model, the  rishons  are  defined  as  quantum states  transforming  
like  the  fundamental (and  its  conjugate) representations of the group 
$ SU_3(H) \times SU_3(C) \times  SU_3(R)$.  
We  first introduce the  rishons  and  their  quantum  numbers $ G=\pm 1/3$  
for hypercolor,  $ Y=\pm 1/3$   for color and  $  Z=\pm 1/3$  for
``R-color", respectively. The fundamental representation  of $ SU_3(H)$ is 
characterized by  the $ G=1/3 $ quantum  number (G=-1/3  for  the 3 conjugate 
representation) while the fundamental representation of $SU_3(C)$ is 
characterized by the Y=1/3 quantum number and  that  of $SU_3(R)$ is 
characterized by the  Z=1/3. The number of electric charge is represented as
\begin{equation}
            Q = {1\over 2} (Y + G)
\label{eq:charge}
\end{equation}
We represent rishon as $R(\alpha i;\lambda)$ where $\alpha$ denotes the 
hypercolor and i the color, $\lambda$ the R-color indices of  the  state.  
The  internal quantum number of $R_{\alpha}$ state is  G=1/3  and  represent  
the  rishon state while  $R^{\alpha}$ is G=-1/3  and  anti-rishon  state.  
Similarly, the quantum number of $R_i$ is Y=1/3 and $R_{\lambda}$ is Z=1/3 etc. 
The  $R^{\alpha}$ state with Z=-1/3 is represented by  symbol $\tilde R $, and
called as ``pre-generation state". 
The correspondence to the T and V states is given as 
\begin{eqnarray}
      R_{\alpha i} = T_{\alpha i}, R^{\alpha i} = \bar{T}^{\alpha i},
      R_{\alpha}^i = V_{\alpha}^i, R^{\alpha}_i = \bar{V}^{\alpha}_i       
\label{eq:RTV}
\end{eqnarray}

 We  now  contract three rishon state on their hypercolor indices, and 
getting a singlet in hypercolor:
\begin{eqnarray}
       \Psi(ijk;\lambda\mu\nu) 
       = \sum_{\alpha\beta\gamma}\vert R_{\alpha}(i\lambda)R_{\beta}(j\mu)
       R_{\gamma}(k\nu)\rangle\epsilon^{\alpha\beta\gamma},
\label{eq:contr1}
\end{eqnarray}
\begin{eqnarray}
       \bar{\Psi}(ijk;\lambda\mu\nu) 
       = \sum_{\alpha\beta\gamma}\langle R^{\alpha}(i\lambda)R^{\beta}(j\mu)
         R^{\gamma}(k\nu)\vert\epsilon_{\alpha\beta\gamma},
\label{eq:contra2}
\end{eqnarray}
where suffix $ijk$ and $\lambda\mu\nu$  represent  upper  or  lower  ones.  
The hypercolor quantum number G of such an object  $\psi(ijk;\lambda\mu\nu)$
will be G=1, and that of  $\bar{\psi}(ijk;\lambda\mu\nu)$ is G=-1. 
Contracting with respect to R-color indices, we obtain
\begin{eqnarray}
         \Psi_{\sigma }(ijk) 
       = \sum_{\lambda\mu\nu\rho}{\Psi(ijk)_{\lambda\mu}}^{\nu}
         \epsilon^{\lambda\mu\rho}\epsilon_{\nu\rho\sigma}  
\label{eq:contr11}
\end{eqnarray}
\begin{eqnarray}
         \bar{\Psi}_{\sigma }(ijk) 
       = \sum_{\lambda\mu\nu\rho}{\bar{\Psi}(ijk)_{\lambda\mu}}^{\nu}
         \epsilon^{\lambda\mu\rho}\epsilon_{\nu\rho\sigma}  
\label{eq:contr22}
\end{eqnarray}

where  the  suffix $\sigma$ represent the generation label.  The contraction 
of color indices leads to color triplets  having  the internal quantum number 
of the U type quarks together with generation label $\sigma$.
\begin{eqnarray}
   U_{\sigma ,m} &=&\sum_{ijkl}{\Psi_{\sigma ,ij}}^k
         \epsilon^{ijl}\epsilon_{klm}\cr
                &=&\sum_{ijkl}\sum_{\lambda\mu\nu\rho}\sum_{\alpha\beta\gamma}
                  \vert R_{\alpha i\lambda}R_{\beta j\mu}R_{\gamma}^{k\nu}\rangle
                  \epsilon^{ijl}\epsilon_{klm}\epsilon^{\alpha\beta\gamma}
                  \epsilon^{\lambda\mu\rho}\epsilon_{\nu\sigma\rho}\cr
                &=&\sum_{ijkl}\sum_{\lambda\mu\nu\rho}\sum_{\alpha\beta\gamma}
                  \vert T_{\alpha i\lambda}T_{\beta j\mu}V_{\gamma}^{k\nu}\rangle
                  \epsilon^{ijl}\epsilon_{klm}\epsilon^{\alpha\beta\gamma}
                  \epsilon^{\lambda\mu\rho}\epsilon_{\nu\sigma\rho} 
\label{eq:U}
\end{eqnarray}
where m represent the color label. The  quantum  number  of  this state is 
given as G=1, and Y=1/3, or equivalently Q=2/3, and Z=1/3.
The configuration of D type quarks is also given by
\begin{eqnarray}
   D_{\sigma ,m} &=& \sum_{ijkl}{{\bar{\Psi}}_{\sigma ,ij}}^{~~~~k}
                  \epsilon^{ijl}\epsilon_{klm}\cr
                &=& \sum_{ijkl}\sum_{\lambda\mu\nu\rho}\sum_{\alpha\beta\gamma}
                  \langle R^{\alpha}_{\sigma \lambda}R^{\beta}_{j\mu}R^{\gamma k\nu}\vert
                  \epsilon^{ijl}\epsilon_{klm}\epsilon_{\alpha\beta\gamma}
                  \epsilon^{\lambda\mu\rho}\epsilon_{\nu\sigma\rho}\cr
                &=& \sum_{ijkl}\sum_{\lambda\mu\nu\rho}\sum_{\alpha\beta\gamma}
                  \langle \bar{V}^{\alpha}_{\sigma \lambda}\bar{V}^{\beta}_{j\mu}
                  \bar{T}^{\gamma k\nu}\vert
                  \epsilon^{ijl}\epsilon_{klm}\epsilon_{\alpha\beta\gamma}
                  \epsilon^{\lambda\mu\rho}\epsilon_{\nu\sigma\rho}
\label{eq:D}
\end{eqnarray}

where G=-1, Y=1/3 ,Q=-1/3 , and Z=1/3.
  
  The singlet  in  hypercolor  and  color  corresponds  to lepton state,
\begin{eqnarray}
  {\ell}_{\sigma } &=& \sum_{ijk}{{\bar{\Psi}}_{\sigma }}^{~ijk}\epsilon_{ijk}\cr
                  &=& \sum_{ijk}\sum_{\lambda\mu\nu\rho}\sum_{\alpha\beta\gamma}
                   \langle \bar{T}^{\alpha,i}_{\lambda}\bar{T}^{\beta,j}_{\mu}
                   \bar{T}^{\gamma k\nu}\vert
                   \epsilon_{ijk}\epsilon_{\alpha\beta\gamma}
                   \epsilon^{\lambda\mu\rho}\epsilon_{\nu\rho\sigma}
\label{eq:L}
\end{eqnarray}
 
with G=-1,Y=-1 ,Q=-1, and Z=1/3.
   
   Similarly, the configuration on  neutrino with generation label $\sigma$
is given by
\begin{eqnarray}
   {\nu}_{\sigma } &=& \sum_{ijk}{{\Psi}_{\sigma }}^{ijk}\epsilon_{ijk}\cr
                  &=& \sum_{ijk}\sum_{\lambda\mu\nu\rho}\sum_{\alpha\beta\gamma}
                   \vert {{V}_{\alpha}^i}_{\lambda}{{V}_{\beta}^{j}}_{\mu}
                   {V}_{\gamma}^{k\nu}\rangle
                   \epsilon_{ijk}\epsilon^{\alpha\beta\gamma}
                   \epsilon^{\lambda\mu\rho}\epsilon_{\nu\rho\sigma}
\label{eq:Nu}
\end{eqnarray}
where G=1, Y=-1, Q=0 , and Z=1/3.

   Thus, in the framework of geometrical model\cite{rf:Geometrical}, the generation 
label can be introduced without any ambiguity\cite{rf:Dyna-R}.

\section{Structure of ``hot-sector generations"}

        In our model, the 3 generation structure of ``cold-sector" generations is  
represented  by $\Psi_{\lambda,\mu}^{\nu}$ representation of $SU_3(R)$ group. It 
should be noted that  there appears further configurations $\Psi_{\lambda,\mu,\nu}$,
$\Psi_{\lambda}^{\mu,\nu}$ and $\Psi^{\lambda,\mu,\nu}$. What is meant by the 
existence  of  these  configurations?   The  most  natural interpretation is to 
identify them to the hot-sector generations. That is, there  are  3  generations 
in hot-sector, which is represented by $\Psi_{\lambda}^{\mu,\nu}$. Further, there 
is the  other  configuration , which is represented as $\Psi^{\lambda,\mu,\nu}$. 
This will mean  the  existence of further one hot generation. That is,  our  model  
suggests  the following generation structure:

\vspace{0.25cm}
   $\Psi_{\lambda,\mu,\nu}~~~   \rm{frozen~sector}~~~~~~~~~~~~~~~~~~~~~~ Z=1$
\vspace{0.15cm}

   $\Psi_{\lambda,\mu}^{\nu}~~~~~ \rm{3~cold~sector~generations}~~~~~~   Z=1/3$
\vspace{0.15cm}

   $\Psi_{\lambda}^{\mu,\nu}~~~~~  \rm{3~hot~sector~generations}~~~~~~~  Z=-1/3$ 
\vspace{0.15cm}

   $\Psi^{\lambda,\mu,\nu}~~~    \rm{1~hot~sector~generation}~~~~~~~~   Z=-1$
\vspace{0.25cm}

The $\Psi_{\lambda,\mu,\nu}$ configuration which contains no pre-generation state
${\rm \tilde R}$ 
should be interpreted that its sector has been frozen by some reason. 
The $\Psi_{\lambda}^{\mu,\nu}$ and  
$\Psi^{\lambda,\mu,\nu}$ configurations represent the hot-sector generations. 
That is, {\bf 3 + 1}  structure of hot-sector generations is expected in our model. 
   Then, what is  meant  by  the  ``hot-sector"  and  ``frozen sector" generations? 
We will stand on  the  view-point  that  the rishon system is  the  one  beyond  
the  ordinary  quantum  field theory, and we have treated only classification 
symmetry  without treating the details of dynamics.  The new  dynamics may be 
related to the quantum field theory with a specific structure and principle, 
or further may be beyond the quantum theory though it seems to be extremely 
applicable. In the present stage, 
however, it is difficult to find out yet to be  known  new dynamics in the 
complete form. 
It is important to note that the new dynamics  should  be the  one  
to  lead  to  the  standard  model  effectively  in  an appropriate 
energy region. From this view-point,  the  fruits  of field theoretical 
approach to sub-system should  be  remarked\cite{rf:Pati-Salam}.
 Especially, it is 
known that the possession of a certain kind  of symmetry, ie. chiral 
symmetry and/or supersymmetry, in the  gauge theory of composite
particle formation leads to  the  realization of the light fermion. 
Some models based  on  this  mechanism  are proposed,  which  predict  
the  existence  of   heavy   eccentric particles\cite{rf:Akama}. It is probable 
that the existence of cold and  hot-sector generations in our model 
is founded by making use of  such mechanism.
The  constitution   of   theory   containing   hot-sector generation 
will be forced to take a form  of  mosaic  of  quantum field theory, 
because that generations are expected to be beyond the ordinary quantum 
field theory. In such a  practice,  the  meaning  of  the frozen-sector may 
be also clarified.  It  is  probable  that  the frozen sector is 
understood as ghost in a space  with  indefinite metric, or it does 
not  form  the  bound  state  in  the  present vacuum, though it 
appears in the specific vacuum such as in  early universe as  bound 
states, etc. As a step to approach to these problems, we will examine
a possible model of the hot-sector generations in the framework of 
present field theory.

\section{Neutrino mass in hot-sector generations}
\indent
 Our model predict possible existence of hot-sector generations with
${\bf 3 + 1}$ structure. However, the result of experiment of  
$Z \rightarrow \overline{\nu_{\ell}}\nu_{\ell}$ 
shows that the number of neutrinos concerning to this process is
just 3 \cite{rf:Gener3}. That is, the number of neutrinos with mass 
below $M_Z/2$ is 
restricted to 3. Then, the neutrino masses of
possible hot-sector generations should be above $M_Z/2$, so far as
the same simple generation structure as cold-sector is 
maintained\cite{rf:no-neutral}.
 
 As is well known, the smallness of ordinary neutrino mass 
is nicely explained by the see-saw mechanism. If this mechanism is realized 
in the neutrinos of cold-sector generations, it is natural to suppose
that a certain kind of see-saw mechanism is also realized in some neutrinos 
belonging to
the hot-sector generations. What mechanism to satisfy the neutrino mass bound
appears in that case?

As a basis for the construction of our scheme, let
us consider the D-M (Dirac-Majorana) mass term 
\cite{rf:GRSY}-\cite{rf:Bilen} in the simplest case of
one generation labeled by the generation subscript $\sigma$.
We have

\begin{eqnarray}
{\cal{L}}^{D-M}& = &
-\frac{1}{2} m_{\sigma L} \overline{(\nu_{\sigma L})^c} \nu_{\sigma  L}
-m_{\sigma  D}\bar{\nu}_{\sigma R}\nu_{\sigma  L}
\nonumber\\
&   & -\frac{1}{2} m_{\sigma R} \bar{\nu}_{\sigma R} (\nu_{\sigma R})^c~~+~~h.c.
                  \nonumber\\
               & = &-\frac{1}{2}
\overline{{(\nu_{\sigma L})^c} \choose {\nu_{\sigma  R}}}
{}~M~{{\nu_{\sigma L}} \choose {\nu_{\sigma R}^c}}
{}~~+~~h.c.
\label{eq:DM}
\end{eqnarray}

\noindent Here
\beq
M = \left( \begin{array}{cc}
m_{\sigma L} & m_{\sigma D} \\
m_{\sigma D} & m_{\sigma R}
\end{array} \right),
\label{eq:MATm}
\end{equation}

\noindent where $m_{\sigma L}, m_{\sigma D}, m_{\sigma R} $ are parameters. For a symmetrical 
matrix $M$ we have
\beq
M=UmU^{\dag},
\label{eq:DIA}
\end{equation}

\noindent where $U^{\dag}U=1 $ and $m_{jk}=m_j \delta_{jk} $. From 
Eqs.~(\ref{eq:DM}) and
(\ref{eq:DIA}) we have

\beq
{\cal{L}}^{\rm{D-M}}=-\frac{1}{2} \sum_{\alpha=1}^{2}
 m_{\sigma {\alpha}}{\bar{\chi}}_{{\sigma\alpha}} \chi_{\sigma {\alpha}} ,
\label{eq:dia2}
\end{equation}

\noindent where

\begin{eqnarray}
\nu_{\sigma L}~~  =~~~{\cos{\theta_\sigma}}\chi_{\sigma 1L} & + & {\sin{\theta_\sigma}}\chi_{\sigma 2L},
\nonumber\\
(\nu_{\sigma R})^c  =  {-\sin{\theta_\sigma}}\chi_{\sigma 1L} & + & {\cos{\theta_\sigma}}\chi_{\sigma 2L}.
\label{eq:MIX}
\end{eqnarray}

\noindent Here $\chi_{\sigma 1} $ and $\chi_{\sigma 2} $ are fields of Majorana
 neutrinos with masses $ m_{\sigma s}~(a~~``small"~ mass)$, $m_{\sigma B}~(a~~``Big"~ mass) $, 
 respectively.
  The masses $m_{\sigma s}$ and $m_{\sigma B} $ and the mixing angle
$\theta_\sigma $ are connected to the parameters $m_{\sigma L}, m_{\sigma D} $ 
and $m_{\sigma R} $ by the relations

\begin{eqnarray}
m_{\sigma s}~ & = & \frac{1}{2}~ {\left|{m_{\sigma R}~ +~m_{\sigma L}~-~a_\sigma} \right|},
\nonumber\\
m_{\sigma B} & = & \frac{1}{2}~ {\left|{m_{\sigma R}~ +~m_{\sigma L}~+~a_\sigma} \right|},
\nonumber\\
\sin{2\theta_\sigma} & = & \frac{2m_{\sigma D}}{a_\sigma},~~~~\cos{2\theta_\sigma},
=\frac{m_{\sigma R}-m_{\sigma L}}{a_\sigma},
\label{eq:MIX2}
\end{eqnarray}
\noindent where
\beq
a_\sigma=\sqrt{(m_{\sigma R}-m_{\sigma L})^2~+~4{m_{\sigma D}^2}}.
\label{eq:a}
\end{equation}
It should be noted that the relations Eq.~(\ref{eq:MIX2}) are exact.

\subsection{\it Heavy neutrinos in hot-sector generation}
 
Let us assume now that
\beq
m_{\sigma L}=m_{\sigma 0},~~~ m_{\sigma D}\simeq{ m_{\sigma F}},~~~m_{\sigma R}\gg{ m_{\sigma F}} ,
\label{eq:mass-sb1}
\end{equation}

\noindent where $m_{\sigma F} $ is a typical mass of the leptons and quarks of
the generation labeled by the subscript $\sigma$.
Here,   $m_{\sigma 0}$ should have an appropriate value above $M_Z$.
If we assume, as a proto type of typical case, that $m_{\sigma B}\simeq {10^{19}} $ GeV 
(Planck mass) and $m_{\sigma 0} = 100GeV$, then we see that 
\beq
m_{\sigma s}\simeq 100GeV,~~m_{\sigma B}\simeq {10^{19}} GeV.
\label{eq:mass-estim}
\end{equation}

Are the heavy neutrinos stable? They may decay into particles in cold sector
through a very small mixing of hot-sector generations with cold sector ones, 
or through the interaction of heavy gauge bosons appearing in GUTs. 

\indent

\subsection{\it See-saw mechanism in cold-sector generations} 

\indent

Instead of Eq.~(\ref{eq:mass-sb1}), if we assume\cite{rf:GRSY}-\cite{rf:Bilen}
\beq
m_{\sigma L}= 0,~ m_{\sigma D}\simeq{ m_{\sigma F}},
~m_{\sigma R}\gg{ m_{\sigma F}},
\label{eq:mass-see}
\end{equation}
it leads to well known see-saw mechanism
\beq
m_{\sigma s}\simeq\frac{m_{\sigma F}^2}{m_{\sigma R}},~~m_{\sigma B}\simeq{m_{\sigma R}},
{}~~\theta_i\simeq{\frac{m_{\sigma D}}{m_{\sigma R}}}
\label{eq:mass-sb2}
\end{equation}
\noindent
 Thus, if the conditions Eq.~(\ref{eq:mass-see})  are satisfied,
the particles with definite masses are split to a very light Majorana
neutrino with mass
$m_{\sigma s} \ll m_{\sigma F} $ and a very  heavy Majorana particle with mass
$m_{\sigma B}\simeq m_{\sigma R} $. The current neutrino field $\nu_{\sigma L} $ practically
coincides with
 $\chi_{\sigma 1L} $ and $ \chi_{\sigma 2}\simeq {\nu_{\sigma R}~+~(\nu_{\sigma R})^c}$,
because $\theta_i$ is extremely small.
That is, we have assumed such scheme that in
D-M mass term Dirac masses  are of order of usual fermion masses, 
the right-handed Majorana masses, responsible for lepton numbers
violation, are  extremely large  and the left-handed 
Majorana masses are equal zero. In such a scheme  neutrinos are Majorana
particles with masses much smaller than masses of the other fermions.
 
\vspace{0.5cm}

\section{Discussion}
   In this paper, we have proposed a  model of realization of 
hot-sector generation. In our model, the neutrino mass of hot-sector 
generations is realized by a certain kind of see-saw mechanism, in
which Majorana mass term of $m_{\sigma L} \overline{(\nu_{\sigma L})^c} \nu_{\sigma L}$ 
type appears.

Our model is based on a schematical formulation of rishon model, 
where the existence of 3-generation structure of cold-sector is naturally explained.
This schematical model should be supposed to be concerning to the sub-structure 
behind the GUTs structure of leptons and quarks.
Though our model can explain the 3-generation structure, it can not explain so
sufficiently why the mass of top quark is so heavy. It is reduced to badly 
broken symmetry caused by yet to be known some mechanism. Natural explanation of the 
large mass of it is further problem. 
Further, precise decision of neutrino mass and oscillation pattern in the lepton sector
will light on the related problems\cite{rf:SK-Atmospher}.

It should be emphasized that almost all quantum numbers including
lepton and quark numbers are not conserved in GUTs. 
The rishon model is just the one based on the most fundamental electric charge, 
which is exactly conserved
in GUTs. In this sense, the rishon model is very remarkable model.
Further, it is probable some of these ``particles"
in hot-sector generations are the ones beyond ordinary particle picture.
It is not yet known how behave these particles.
The problem of the upper bound of flavor number in ordinary field theory should be
examined in this context\cite{rf:GW}.

Finally, if new event concerning to new particles is discovered, we should 
examine the possibility that it is the one belonging to the hot-sector generations,
together with one in GUTs or super-symmetric GUTs.

\pagebreak



\begin{thebibliography}{99}
\bibitem{rf:quark0}
It should be noted that the quark had  been  treated,  in  the first 
paper proposing it, as if  it  is  ``purely  mathematical  entities" 
in order to avoid some  difficulties.  In  fact, the problem why does 
qqq form a bound state  had  been  completely  clarified after the 
discovery of QCD.
\bibitem{rf:no-neutral}
In the early stage of electroweak theory, some electroweak models with 
no neutral current had been proposed. They are based on  group such as
$SU(3) \times U(1)$, and have ordinary charged weak current. We don't
treat in present these cases because these models seem to not satisfy 
the realization of simple generation structure.
\end{thebibliography}
\end{document}